\documentclass[12pt]{iopart}
\usepackage{iopams}
\usepackage{braket}
\usepackage{cases}
\usepackage[dvipdfmx]{graphicx}

\newcommand{\beq}{\begin{equation}}
\newcommand{\eeq}{\end{equation}}
\newcommand{\beqn}{\begin{eqnarray}}
\newcommand{\eeqn}{\end{eqnarray}}

\begin{document}
\title{Novel symmetry-broken phase in a driven cavity system in the thermodynamic limit}
\author{T Shirai$^1$, T Mori$^1$, S Miyashita$^{1,2}$}
\address{%
$^1$ Department of Physics, Graduate School of Science,
The University of Tokyo, 7-3-1 Hongo, Bunkyo-Ku, Tokyo 113-8656, Japan
}
\address{%
$^2$ CREST, JST, 4-1-8 Honcho Kawaguchi, Saitama 332-0012, Japan
}%

\ead{\mailto{shirai@spin.phys.s.u-tokyo.ac.jp}, \mailto{mori@spin.phys.s.u-tokyo.ac.jp}, \mailto{miya@spin.phys.s.u-tokyo.ac.jp}}           %  \\

\begin{abstract}
We study non-equilibrium stationary states of a cavity system consisting of many atoms interacting with a quantized cavity field mode,
under a driving field in a dissipative environment.
We derive a quantum master equation which is suitable for treating systems with a strong driving field and a strong atom-photon interaction.
We do this by making use of the fact that the mean-field dynamics are exact in the thermodynamic limit
thanks to a uniform coupling between atoms and photons.
We find ordered states with symmetry-broken components of the photon field and atomic excitation driven by the external field.
The mechanism by which these ordered states arise is discussed from the viewpoint of the quantum interference effect.
\end{abstract}

\vspace{2pc}

\submitto{\JPB}

\maketitle

\section{Introduction}\label{Sec_one}
The effects of interaction between photons and atoms have been studied for a long time.
A cavity is introduced to enhance the interaction by confining photons in a finite region with mirrors.
Such a system is described by a two-level atom coupled with a single-mode photon field which represents the photons in the cavity.
This system is described by the Jaynes-Cummings model with the rotating-wave approximation (RWA)~\cite{Jaynes1963,Shore_rev1993}.
When many atoms are included, the model is extended to the Tavis-Cummings model~\cite{Tavis1968}.
The coupling effects have been experimentally observed as cavity ringing phenomena~\cite{Kaluzny1983,Raizen1989} and vacuum Rabi field splitting~\cite{ZhuY1990}.
Recently, this coupling has attracted much attention as a possible method to control quantum information~\cite{Chiorescu2010,Schuster2010,KuboY2010,Blencowe2010,Amsuss2011,ZhuX2011,KuboY2011}.

There have also been many studies on cooperative phenomena since Dicke noted the importance of the uniform coupling between many atoms and photons~\cite{Dicke1954}.
In the strong coupling (SC) region where the coupling strength is as strong as the dissipation constants, but still much smaller than the energy scale of the two-level atoms and photons,
an optical bistability appears under a driving external field~\cite{wolf_text, Drummond1981}.
In this transition, the stationary state changes discontinuously as a function of the strength and the frequency of the external field, which has been observed in experiments~\cite{Rempe1991, Gripp1996}.
When the atom-photon coupling becomes much stronger and the coupling strength is comparable with the energy of two-level atoms and photons, we call such a region the `ultra-strong coupling (USC) region'.
In this region, the so-called Dicke transition~\cite{Hepp1973_ann, Wang1973, Hepp1973} occurs at a critical value of the atom-photon coupling strength.
In the ordered phase, the photon number of the ground state is not zero and the dipole moment of atoms is spontaneously polarized.
This phase is called the `superradiant phase'~\cite{Hepp1973_ann}.
With recent experimental developments in studies on many-body systems,
it becomes possible to realize the USC region.
For instance, a phase transition corresponding to the Dicke transition has been studied in a cold atom system in an optical cavity~\cite{Baumann2010},
and phenomena induced by parametric resonance were proposed theoretically in this system~\cite{Bastidas2012,Vacanti2012}.
Furthermore, the USC region is also realized in other systems, e.g. semiconductor cavities~\cite{Gunter2009,Anappara2009,Todorov2010,Geiser2012} and circuit QED systems~\cite{Niemczyk2010,Fedorov2010}.
Circuit QED systems with multiple qubits have not yet been realized experimentally,
but the number of qubits is expected to increase in the coming years~\cite{Garraway2011}.

In the present paper, we study the long-time asymptotic states of the Dicke model in the USC region under a strong driving external field.
For this, we need to extend the master equation conventionally used in studies of optical bistability.
In~\cite{Murao1995,Beaudoin2011}, the dissipation effect for a single two-level atom coupled with a cavity photon field was studied,
and the importance of incorporating the effects of the interaction between photons and atoms into the dissipation effect was pointed out.
However, for cooperative phenomena, this effect has not yet been investigated.
Thus, the types of cooperative phenomena existing in this region are not known.
Our study provides an indication as to the phenomena, and this is to be realized in experiments.
By using the master equation derived in this paper, under a strong driving field we find a novel kind of symmetry-breaking phenomenon which is different from the Dicke transition.
In order to make clear what is essential for the appearance of this symmetry-broken state, we also study the Tavis-Cummings model for comparison.
It is known that the Tavis-Cummings model is not adequate as a model of the cavity system when the interaction is strong,
but optical bistability and a Dicke transition occur even if we adopt the RWA.
Thus, it is important to check the effect of the RWA in the USC region under a strong driving field.
We find that the two models exhibit qualitatively different stationary states;
that is, the novel symmetry breaking does not appear in the case of the Tavis-Cummings model.
This observation shows that the RWA has a significant effect in this region.

We give a physical interpretation for the novel symmetry breaking in the Dicke model as a synergistic phenomenon of a quantum interference effect induced by a strong driving external field and the strong interaction effect.
We also give an explanation of why the two models show qualitatively different properties in this region.

The paper consists of the following sections:
In section~\ref{Sec_two}, the driven Dicke model is explained.
In section~\ref{Sec_three}, we derive a new master equation for the USC region under a strong driving external field in the thermodynamic limit.
In section~\ref{Sec_four}, we present the novel symmetry-breaking phenomenon induced by a strong driving external field,
and give an physical interpretation of this phenomenon.
In section~\ref{Sec_five}, we summarize the paper.

\section{The model of the cavity system}\label{Sec_two}
We study cooperative phenomena in the cavity system described by the Dicke model~\cite{Dicke1954},
i.e. a group of two-level atoms coupled with the single-mode photon field whose Hamiltonian is given by
\beq
\fl H_{\rm S}(t)=\omega_{\rm p} a^{\dagger} a+\sum_{j=1}^N \omega_{\rm a} S_j^z +\sum_{j=1}^N \frac{2g}{\sqrt{N}} S_j^x (a+a^{\dagger})+2\sqrt{N}\xi \cos ( \omega_{\rm e} t ) (a+a^{\dagger}).\label{system}
\eeq
The first term describes the single-mode photon field whose energy is proportional to the frequency $\omega_{\rm p}$.
Throughout the paper, we set $\hbar=1$.
The annihilation and creation operators of the photon mode are denoted by $a$ and $a^{\dagger}$, respectively.
They satisfy the commutation relation of bosonic operators, $[a,a^{\dagger}]=1$.
The second term denotes an ensemble of the two-level atoms whose energy gaps are proportional to the atomic frequency $\omega_{\rm a}$.
They are described by spin operators $\bi{S}_j=(S_j^x, S_j^y, S_j^z)=(\sigma^x/2, \sigma^y/2, \sigma^z/2)$
where $\{ \sigma^{\alpha} \}_{\alpha=\{ x, y, z\}}$ are the Pauli matrices.
The third term gives the interaction between photons and two-level atoms, and $g$ is the coupling strength.
The last term describes the driving external field with amplitude $\xi$ and frequency $\omega_{\rm e}$.
In the present work, we consider the thermodynamic limit, $N \rightarrow \infty$.
In this limit, the expectation values of $a^{\dagger}a$ and $\sum_{j=1}^N \bi{S}_j$ are of $\Or(N)$,
and therefore, in equation (\ref{system}), we adopt the rescaled parameters for the atom-photon coupling and the amplitude of the driving external field.

The Tavis-Cummings model~\cite{Tavis1968} is obtained by applying the RWA to the Dicke model,
\beq
\sum_{j=1}^N \frac{2g}{\sqrt{N}} S_j^x (a+a^{\dagger}) \rightarrow \sum_{j=1}^N \frac{g}{\sqrt{N}} (S_j^+ a +S_j^- a^{\dagger}).
\eeq
It is known that the two models exhibit a Dicke transition when $g$ is larger than a critical value.
In the Dicke model the critical value is given by $g_{\rm Dicke}=\sqrt{\omega_{\rm a} \omega_{\rm p}}/2$~\cite{Hepp1973},
while in the Tavis-Cummings model the critical value is given by $g_{\rm TC}=\sqrt{\omega_{\rm a}\omega_{\rm p}}$~\cite{Hepp1973_ann, Wang1973} at zero temperature.

It should be noted that the symmetries of the two models are different;
The Dicke model has a $Z(2)$ symmetry, while the Tavis-Cummings model has a $U(1)$ symmetry.
This difference causes significant effects on the stationary states in the USC region under a strong driving field, as we will see later.

\section{The dressed Lindblad equation}\label{Sec_three}
In this section, we derive a master equation to study properties in the region where the interaction and the driving field are strong.
Here, the derivation is given only for the Dicke model, but the formalism of the master equation is also straightforward for the Tavis-Cummings model.

\subsection{Thermal bath}
The master equation describes the time evolution of the system (S) in a dissipative environment.
In order to take the dissipation effect into account, we prepare a thermal bath (B) in contact with the system.
For simplicity, in this study we assume that the temperature of the thermal bath is zero.
The Hamiltonian of the total system is given by
\beqn
H_{\rm T} (t)=H_{\rm S}(t)+H_{\rm B}+\lambda H_{\rm I},
\eeqn
where $H_{\rm S}(t)$ and $H_{\rm B}$ are the Hamiltonians of the system and the thermal bath, respectively,
and $H_{\rm I}$ is the interaction between them.
Here, we assume that both the photons and atoms in the cavity system interact independently with different thermal baths;
$H_{\rm B}=H_{\rm B}^{\rm P}+H_{\rm B, Local}^{\rm A}+H_{\rm B, Global}^{\rm A}$
and $H_{\rm I}=H_{\rm I}^{\rm P}+H_{\rm I, Local}^{\rm A}+H_{\rm I, Global}^{\rm A}$.
For the cavity photons, we adopt a free boson bath given by
\beq
H_{\rm B}^{\rm P}=\sum_{\alpha} \omega_{\alpha} A_{{\rm p}, \alpha}^{\dagger} A_{{\rm p}, \alpha},
\eeq
where $\{ A_{{\rm p}, \alpha} \}$ and $\{ A_{{\rm p}, \alpha}^{\dagger} \}$ are the annihilation and the creation bosonic operators, respectively.
The bath is in contact with the cavity photons in a bilinear form:
\beq
H_{\rm I}^{\rm P}=\sum_{\alpha} (k_{\alpha} A_{{\rm p}, \alpha}+k_{\alpha}^* A_{{\rm p}, \alpha}^{\dagger})(a+a^{\dagger}).
\eeq
For the atoms in the cavity,
it should be noted that we need to consider two types of interactions~\cite{Carmichael_text};
one of them describes the interaction between each atom and its own thermal bath (local coupling bath),
\beqn
\fl H_{\rm B, Local}^{\rm A}+\lambda H_{\rm I,Local}^{\rm A}=\sum_{j=1}^N \sum_{\alpha} \omega_{j, \alpha} A_{{\rm L}j, \alpha}^{\dagger} A_{{\rm L}j, \alpha}\nonumber\\
\qquad\quad+\lambda \sum_{j=1}^N \sum_{\alpha} ( k_{j, \alpha} A_{{\rm L}j, \alpha} + k_{j, \alpha}^* A_{{\rm L}j, \alpha}^{\dagger})(S_j^+ +S_j^-),\label{Local}
\eeqn
where $\{ A_{{\rm L}j, \alpha} \}_{j=\{1, \ldots, N\}}$ and $\{ A_{{\rm L}j, \alpha}^{\dagger} \}_{j=\{1, \ldots, N\}}$ are the annihilation and the creation bosonic operators, respectively.
In the other type of interaction, all the atoms couple with a single thermal bath (global coupling bath),
\beqn
\fl H_{\rm B, Global}^{\rm A}+\lambda H_{\rm I, Global}^{\rm A}&=\sum_{\alpha} \omega_{{\rm G}, \alpha} A_{{\rm G}, \alpha}^{\dagger} A_{{\rm G}, \alpha}\nonumber\\
&+\frac{\lambda}{\sqrt{N}} \sum_{\alpha} ( k_{{\rm G}, \alpha} A_{{\rm G}, \alpha} + k_{{\rm G}, \alpha}^* A_{{\rm G}, \alpha}^{\dagger})\sum_{j=1}^N (S_j^+ +S_j^-),\label{Global}
\eeqn
where $\{ A_{{\rm G}, \alpha} \}$ and $\{ A_{{\rm G}, \alpha}^{\dagger} \}$ are the annihilation and the creation bosonic operators for this type of  thermal bath, respectively.
The global coupling bath describes the case in which the modes of the bath, e.g. radiation fields, couple with atoms homogeneously. 
In this case, the total spin, $(\sum_{j=1}^N \bi{S}_j)^2=(\sum_{j=1}^N S_j^x)^2+( \sum_{j=1}^N S_j^y)^2+( \sum_{j=1}^N S_j^z )^2$, is conserved. 

In a real system, the total spin relaxes, and hence we need the local coupling bath.
The dependences of the stationary states on the types of bath coupling will be studied later (section~\ref{bath_type}).

\subsection{Master equation}

The density matrix of the total system $\rho_{\rm T} ( t )$ obeys the Liouville-von Neumann equation,
\beqn
\frac{\rmd \rho_{\rm T} (t)}{\rmd t}=-\rmi [H_{\rm T}, \rho_{\rm T} (t)]. \label{Liouville}
\eeqn
In order to derive the reduced dynamics of the system, we introduce a reduced density matrix,
\beqn
\rho_{\rm S} (t)={\rm Tr}_{\rm Bath} \rho_{\rm T}(t).
\eeqn
The dynamics of the reduced density matrix are given by a quantum master equation
which is obtained by projecting out the bath degrees of freedom in equation \eref{Liouville}~\cite{Nakajima1958, Zwanzig1960, Kubo_text, Breuer_text}.
In order to derive the master equation for the region of strong interaction $g$ and strong driving external field $\xi$, 
it is necessary to take the effects of the interaction and driving field into account.
For this, we need all the eigenvalues and eigenstates of the system.
Such a treatment can be done for systems with small degrees of freedom~\cite{Murao1995, Beaudoin2011, Saito2001, Altintas2013} and also for a harmonic chain~\cite{STM2000}.
However, it is difficult in the present case because the system consists of many degrees of freedom.
To overcome this difficulty, we focus on the fact that for the cavity system, many atoms are uniformly coupled with the single-mode photon field.
With this property, we can approach the many-body problem by using a mean-field (MF) strategy, which gives the correct result in the thermodynamic limit~\cite{Mori2013}.
In the MF treatment, the density matrix of the system is given by the product of the density matrix of the photons and atoms,
\beqn
\rho_{\rm S}(t)=\rho_{\rm p}(t) \otimes \underbrace{\rho_{\rm a}(t)\otimes \cdots \otimes \rho_{\rm a}(t)}_{N}=\rho_{\rm p}(t) \otimes \rho_{\rm a}(t)^{\otimes N},
\eeqn
where $\rho_{\rm p}(t)$ is the density matrix of the photons and $\rho_{\rm a}(t)$ is the density matrix of each atom, respectively.
Here, we assume that all the atoms are described by the same density matrix.
The MF Hamiltonian for the photons is given by
\beqn
H^{\rm MF}_{\rm p}(t)&={\rm Tr}_{\rm atoms} H_{\rm S} (t) \rho_{\rm a}(t)^{\otimes N}\nonumber\\
&=\omega_{\rm p} a^{\dagger} a +2\sqrt{N} \left[ g \langle S^x \rangle +\xi \cos(\omega_{\rm e} t) \right] (a+a^{\dagger})+{\rm const.}\nonumber\\
&=\omega_{\rm p} \tilde{a}^{\dagger} \tilde{a} +{\rm const.},\label{dressed_photon}
\eeqn
where $ \langle \bi{S} \rangle={\rm Tr} \bi{S} \rho_{\rm a}(t)$, and
\beq
\tilde{a}=a+\frac{2\sqrt{N}}{\omega_{\rm p}} \left[ g \braket{S^x}+\xi \cos (\omega_{\rm e} t) \right],
\eeq
which is a bosonic operator for a dressed photon which incorporates the MF of the atoms.
In a similar way, the MF Hamiltonian for one of the atoms is given by
\beqn
H^{\rm MF}_{\rm a}(t)&={\rm Tr}_{\rm photon,\hspace{1mm}other\hspace{1mm} atoms} H_{\rm S} (t) \rho_{\rm p}(t)\otimes \hat{1} \otimes \underbrace{\rho_{\rm a}(t)\otimes \cdots \otimes \rho_{\rm a}(t)}_{N-1}\nonumber\\
\fl &=\omega_{\rm a} S^z +\frac{2g}{\sqrt{N}}(\langle a \rangle + \langle a^{\dagger} \rangle ) S^x +{\rm const.},~\label{atom_Hamiltonian}
\eeqn
where $\langle a \rangle ={\rm Tr}a \rho_{\rm p} (t)$.
Its diagonalized Hamiltonian is
\beqn
H^{\rm MF}_{\rm a}(t)=2\sigma (t) \tilde{S}^z+{\rm const.},\label{dressed_atom}
\eeqn
where $\pm \sigma(t)$ are the eigenvalues of the Hamiltonian for the atom given by
\beqn
\sigma (t)=\left[ \left( \frac{\omega_{\rm a}}{2} \right)^2+\frac{g^2}{N} (\langle a \rangle+ \langle a^{\dagger} \rangle )^2 \right]^{\frac{1}{2}},
\eeqn
and $\tilde{\bi{S}}$ describes the dressed atom which incorporates the MF of the cavity photon field.

For the study of the USC region under a driving field, we use `the dressed Lindblad equation' with a Lindblad form~\cite{Breuer_text, Lindblad1976} in terms of the dressed quantities, $\tilde{a}$ and $\tilde{\bi{S}}$.
\begin{numcases}{}
\frac{\rmd \rho_{\rm p}(t)}{\rmd t}=-\rmi [H_{\rm p}(t), \rho_{\rm p}(t)]+\kappa (2 \tilde{a} \rho_{\rm p} (t) \tilde{a}^{\dagger}-\{ \tilde{a}^{\dagger} \tilde{a}, \rho_{\rm p}(t) \} ), \label{extend_photon}\\
\frac{d \rho_{\rm a}(t)}{dt}=-\rmi [H_{\rm a}(t), \rho_{\rm a}(t)]+\gamma_{\rm L} (t) (2 \tilde{S}^- \rho_{\rm a} (t) \tilde{S}^+ -\{ \tilde{S}^+ \tilde{S}^-, \rho_{\rm a}(t) \} )\nonumber\\
\qquad\qquad +\gamma_{\rm G} (t) (\langle \tilde{S}^- \rangle [ \rho_{\rm a} (t), \tilde{S}^+] + \langle \tilde{S}^+ \rangle [\tilde{S}^-, \rho_{\rm a} (t) ] ),\label{extend_atom}
\end{numcases}
where 
\beqn
\kappa =\pi \lambda^2 \sum_{\alpha} |k_{\alpha}|^2 \delta_{\omega_{\alpha}, \omega_{\rm p}},\\
\gamma_{\rm L} (t)=4\pi \lambda^2 |\bra{\tilde{-}} S^x \ket{\tilde{+}}|^2 \sum_{\alpha} |k_{j, \alpha}|^2 \delta_{\omega_{j, \alpha}, \sigma}=4 |\bra{\tilde{-}} S^x \ket{\tilde{+}}|^2 \gamma_{\rm L},\label{local_coupling_bath}\\
\gamma_{\rm G} (t)=4\pi \lambda^2 |\bra{\tilde{-}} S^x \ket{\tilde{+}}|^2 \sum_{\alpha} |k_{{\rm G}, \alpha}|^2 \delta_{\omega_{{\rm G}, \alpha}, \sigma}=4 |\bra{\tilde{-}} S^x \ket{\tilde{+}}|^2 \gamma_{\rm G}.\label{global_coupling_bath}
\eeqn
Here, $\{ \ket{\tilde{-}} \}$ and $\{ \ket{\tilde{+}} \}$ are eigenenergy states of the dressed atom,
\beq
\tilde{S}^z \ket{\tilde{\pm}}=\pm \frac{1}{2}\ket{\tilde{\pm}}.
\eeq
In this dressed Lindblad equation, the effects of the interaction and the driving external field are incorporated into the dressed quantities $\tilde{a}$ and $\tilde{\bi{S}}$.
In this case, the system indeed relaxes into the true ground state of the Hamiltonian at zero external field ($\xi=0$).
Thus, the present formalism of dissipation terms satisfies the minimal condition for the study of the USC region under a strong driving external field.
The dressed Lindblad equation becomes the `bare Lindblad equation' with a Lindblad form in terms of the bare operators $a$ and $\bi{S}$ for $( g, \xi ) \ll ( \omega_{\rm p}, \omega_{\rm a} )$.
In appendix A, we discuss the applicability of this bare Lindblad equation in which the effects of the atom-photon interaction and the driving field are not incorporated into the dissipation terms.
Because the optical bistability in this system occurs when the interaction strength is of the order of the dissipation constants, $( \gamma_{\rm L}, \gamma_{\rm G}, \kappa ) \approx ( g, \xi ) \ll ( \omega_{\rm a}, \omega_{\rm p} )$~\cite{Drummond1981},
the bare Lindblad equation can be used.
However, the study of the properties of the region where the interaction and the driving field are strong requires the use of the dressed Lindblad equation.
\section{The novel symmetry-broken phase induced by the driving external field}\label{Sec_four}

\subsection{Simulation}
In the present work, we study stationary states for the case $\omega_{\rm a}=\omega_{\rm p}=\omega_{\rm e}=1$ by using the dressed Lindblad equations~(\ref{extend_photon}) and (\ref{extend_atom}).
The time evolution of the photon field,
\beq
\alpha (t)=\lim_{N \rightarrow \infty}\frac{\langle a(t) \rangle}{\sqrt{N}},
\eeq
and the atoms,
\beq
\bi{m}(t)=\lim_{N \rightarrow \infty}\frac{1}{N} \sum_{j=1}^N \langle \bi{S}_j (t) \rangle,
\eeq
is investigated.
We solve the equations of motion by means of the Runge-Kutta method.
We regard the state after a sufficiently long time as the stationary state.
In practice, we study the quantities at large $t$ ($\sim 10000 \pi$) which we believe to be large enough to study the stationary state.

\subsection{Classification of the stationary states and order parameter}
In order to study properties of the stationary states, we classify the stationary states into three phases; i.e., (i) regular oscillating phase, (ii) ordered phase, and (iii) non-periodic phase.
Here, we focus on two kinds of symmetry of the system Hamiltonian $H_{\rm S}(t)$.
One is the discrete time-translation symmetry.
Since the period of the external field is $T_{\rm e}=2\pi/\omega_{\rm e}$, the system Hamiltonian is invariant under the transformation $t \rightarrow t+T_{\rm e}$. 
The other symmetry is related to the unitary operator given by
\beqn
U=\exp \left\{ \rmi \pi \left( a^{\dagger}a+\sum_{j=1}^N S_j^z \right) \right\},
\eeqn
which changes the sign of the operators $Y$ $\in$ $\{$ $a/\sqrt{N}$, $a^{\dagger}/\sqrt{N}$, $\{ S_j^x \}$, $\{ S_j^y \}$ $\}$, i.e., $U^{\dagger} Y U = -Y$.
The system Hamiltonian is then invariant under the unitary transformation with the extra time translation $t\rightarrow t+T_{\rm e}/2$,
\beqn
H_{\rm S} (t)=U^{\dagger} H_{\rm S} \left( t+\frac{T_{\rm e}}{2} \right) U.\label{half_symmetry}
\eeqn

The nature of time evolution in stationary states is qualitatively different depending on whether the two kinds of symmetry are broken or not.
If both symmetries hold, the following relation,
\beqn
\langle Y(t) \rangle=-\langle Y ( t+T_{\rm e}/2 ) \rangle, \label{no_order}
\eeqn
should be satisfied.
In this case, $\langle Y(t) \rangle$ oscillates around the origin with the period $T_{\rm e}$, and we call such a case the `regular oscillating phase'.
If the discrete time-translation symmetry holds while the symmetry~(\ref{half_symmetry}) is broken,
the relation~(\ref{no_order}) is not satisfied.
In this case, $\langle Y(t) \rangle$ oscillates around a non-zero value with the period $T_{\rm e}$.
We call this case the `ordered phase'.
When the discrete time-translation symmetry is broken,
the period of $\langle Y(t) \rangle$ is different from $T_{\rm e}$, and such a state is said to be in the `non-periodic phase'.
In this non-periodic phase, the system shows a long-period oscillation or even chaotic behaviour, in the sense that $\langle Y (t) \rangle$ does not show periodic behaviour at all.
The nature of the non-periodic phase is not studied in detail in the present work.

In order to distinguish the three phases,
we consider $\{ \alpha_{j} \}_{j=0, 1, 2, \cdots}$, which is a sequence of the time average of $\alpha (t)$ over one period $(j T_{\rm e}\leq t \leq (j+1) T_{\rm e})$. 
We define the mean value of $\alpha_{j}$ as
\beqn
\alpha_{\rm order}=\lim_{n\rightarrow \infty}\frac{1}{n} \sum_{j=0}^{n-1} \alpha_{j}=\lim_{T \rightarrow \infty} \frac{1}{T} \int_{0}^{T} \alpha (\tau ) d \tau,\label{mean}
\eeqn
and the fluctuation of $\alpha_j$ as
\beqn
\sigma_{\alpha}=\lim_{n \rightarrow \infty} \sqrt{\frac{1}{n} \sum_{j=0}^{n-1} ( \alpha_j -\alpha_{\rm order} )^2}.\label{deviation}
\eeqn
With these quantities, we characterize the three phases as follows:
\begin{enumerate}
\item[(i)] $regular$ $oscillating$ $phase$: the system oscillates with the period $T_{\rm e}$, and $\alpha (t)$ oscillates around zero;
$\alpha_{\rm order}=0$ and $\sigma_{\alpha}=0$;
\item[(ii)] $ordered$ $phase$: the system oscillates with the period $T_{\rm e}$, and $\alpha (t)$ shows a symmetry breaking;
$\alpha_{\rm order} \neq 0$ and $\sigma_{\alpha}=0$;
\item[(iii)] $non$-$periodic$ $phase$: $\alpha (t)$ does not show a simple oscillation.
In this case, $\sigma_{\alpha} \neq 0$.
\end{enumerate}

\subsection{Symmetry-broken phase}
In the present work, we adopt the stationary state of the system without the external field $(\xi=0)$ for the given parameter set as the initial state for all the simulations.
The external field is then applied $(\xi >0)$, and the system evolves in time.
After a sufficiently long time, the system reaches a stationary state,
and we calculate $\alpha (t)$, from which we obtain the order parameters, $\alpha_{\rm order}$ and $\sigma_{\alpha}$.

\begin{figure}[hbt]
\begin{center}
\begin{tabular}{cccc}
(a)&&(b)\\
&\includegraphics[width=64mm,keepaspectratio,clip]{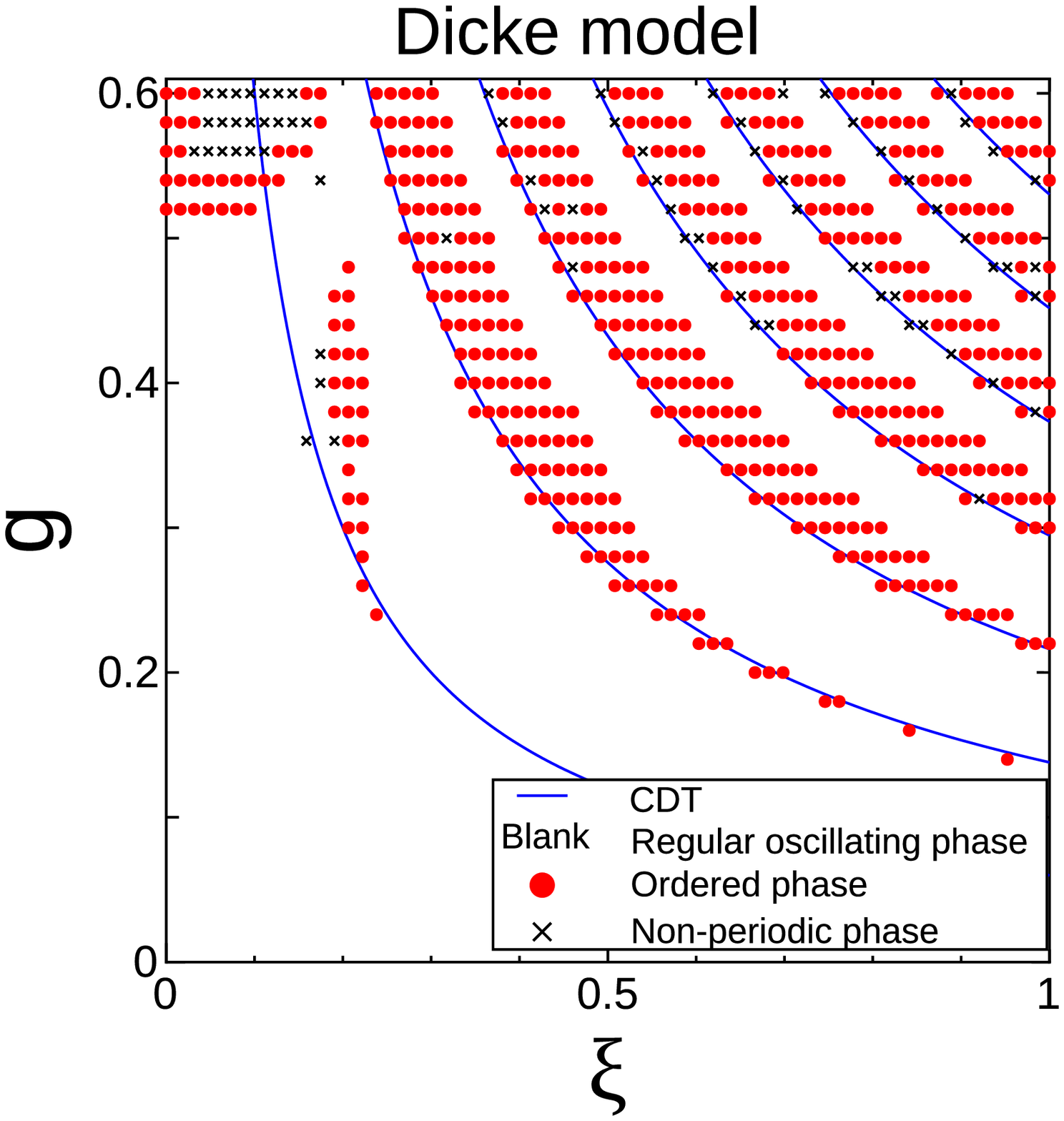}
&&\includegraphics[width=64mm,keepaspectratio,clip]{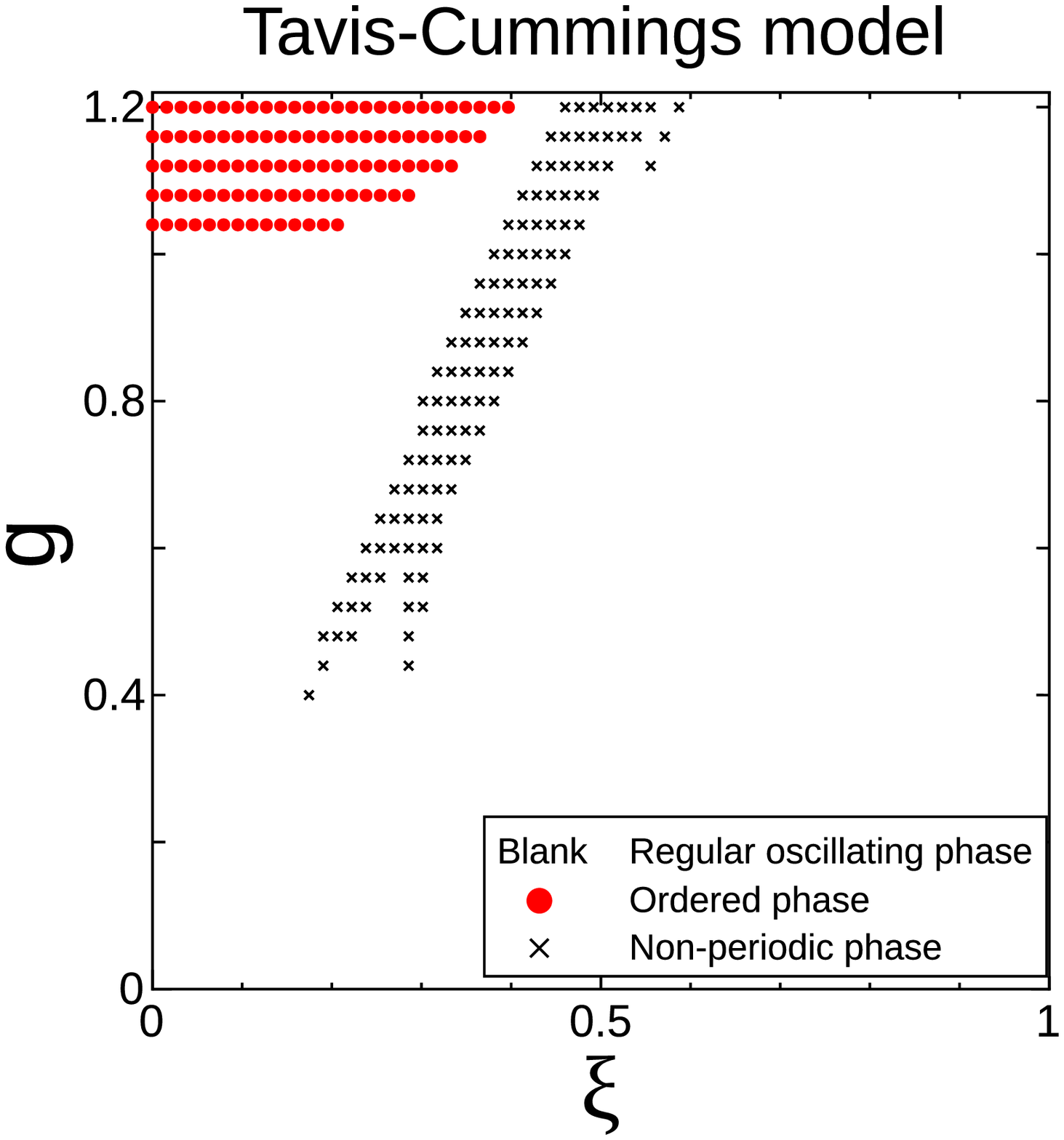}
\end{tabular}
\end{center}
\caption{
Phase diagrams parameterized by the interaction strength $g$ and
the amplitude of the external field $\xi$ of the Dicke model (a) and the Tavis-Cummings model (b)
for the case $\kappa=0.1$ and $\gamma_{\rm L}=0.1$ and $\gamma_{\rm G}=0$.
In figure 1 (a), the positions of CDT are plotted by using blue curves, where $J_0 (4g\xi/\kappa\omega_{\rm e})=0$.
The periodicity of CDT agrees with that of the ordered phase,
which will be discussed in section~\ref{physical_interpretation}.
}
\label{diagram}
\end{figure}

In figure~\ref{diagram}, we present the phase diagram parameterized by $g$ and $\xi$ for the Dicke model and the Tavis-Cummings model.
There, we classify the parameter space $(g, \xi)$ into the following three phases;
blank for the regular oscillating phase, bullets for the ordered phase, and crosses for the non-periodic phase.

In both phase diagrams, the ordered phase at $\xi=0$ originating from the Dicke transition appears when the interaction strength exceeds the critical value ($g_{\rm Dicke}=0.5$ and $g_{\rm TC}=1.0$).
When the external field is applied $(\xi \neq 0)$, the order parameter decreases, and disappears at a certain value of $\xi$.
Therefore we observe that the order of the Dicke transition is destroyed by the driving field.

\begin{figure}
\begin{center}
\includegraphics[width=64mm,keepaspectratio,clip]{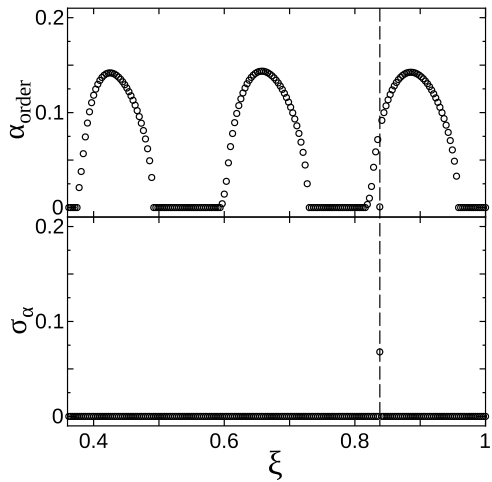}
\end{center}
\caption{
Dependence of the photon order parameter $\alpha_{\rm order}$ (top) and $\sigma_{\alpha}$ (bottom) in the stationary states where the interaction is strong ($g=0.35$) and the driving field is also strong ($0.36 \leq \xi \leq 1$) for the Dicke model.
The ordered phase appears repeatedly as a function of $\xi$.
The dotted vertical line shows that the system is in the non-periodic phase.
The coupling constants are set to be $(\kappa, \gamma_{\rm L}, \gamma_{\rm G})=(0.1, 0.1, 0)$.
}
\label{driven_order_Dicke}
\end{figure}

In the Dicke model (figure~\ref{diagram}(a)), however, besides this ordered phase of the Dicke transition,
a novel type of ordered phase appears at strong driving fields.
In figure~\ref{driven_order_Dicke},
we show the calculated values of $\alpha_{\rm order}$ and $\sigma_{\alpha}$ as a function of $\xi$ for $g=0.35$.
The regions of the ordered phase form a characteristic structure.
For example, they appear repeatedly and form a belt-like structure which extends along a curve given by $g \xi=$ constant, as seen in figure~\ref{diagram}(a).
Since this new phase appears at values of $g$ less than the critical value of the Dicke transition $(g_{\rm Dicke}=0.5)$,
the symmetry-breaking state in this region is not simply originating from the Dicke transition.
The photons and atoms driven by the external field exhibit a spontaneous symmetry-breaking phenomenon.
This phenomenon should be due to the synergistic effects of the interaction and the driving external field.
We do not know of such a phenomenon, and the present observation indicates the existence of a novel type of symmetry-broken state.
Since this phenomenon appears at relatively small values of $g$, we expect that this phase can be realized in experiments.

In contrast, in the Tavis-Cummings model,
this type of ordered phase does not appear at all.
Instead, we find a non-periodic phase in the region where a driving field is strong, which exists along a line $g/\xi=$ constant. 

This observation indicates that the RWA causes qualitatively different phenomena in the region of strong interaction and strong driving field.
The difference is considered to originate from the difference in the symmetry of both models;
that is, the $Z (2)$ symmetry of the Dicke model and the $U (1)$ symmetry of the Tavis-Cummings model.
Significant consequences of the difference between the Dicke model and Tavis-Cummings model have been pointed out for phenomena in the superradiant phase~\cite{Emary2003, Ye2011}.
The present observation of the novel type ordered phase reaffirms the important effects of the symmetry.

\subsection{Physical interpretation of the novel symmetry-broken states}\label{physical_interpretation}
In this subsection, we give a qualitative interpretation of the non-equilibrium phase transition observed in the Dicke model,
and explain why the qualitative differences appear between the Dicke model and the Tavis-Cummings model.
We focus on the effects of the cavity photons on atoms in the region of a strong driving field.
The time evolution of the photons $\alpha (t)$ obtained by the dressed Lindblad equation~(\ref{extend_photon}) is given by
\begin{numcases}{\qquad\qquad\fl
}
\frac{d}{dt}\alpha (t)=(-\rmi \omega_{\rm p}-\kappa )\left[\alpha (t)+\frac{2}{\omega_{\rm p}} (g m^x (t)+\xi \cos (\omega_{\rm e} t))\right] &({\rm Dicke}),\label{photon_eq_Dicke}\\
\frac{d}{dt}\alpha (t)=(-\rmi \omega_{\rm p}-\kappa )\left[\alpha (t)+\frac{1}{\omega_{\rm p}} (g (m^x (t)-\rmi m^y (t))+2\xi \cos (\omega_{\rm e} t))\right] &({\rm T-C}).
\end{numcases}
Under the strong driving field, we can treat the interaction term, which is of the order of $g$, as a perturbation term.
We derive an effective spin model which consists of only the freedom of atoms by using a perturbation method.
As we will see later, the leading-order term describes the quantum interference effect of atoms under a classical photon field,
which is the key ingredient of the present non-equilibrium phase transition.
The next-to-leading-order term describes the long-range interaction between atoms, which leads to the transition as a cooperative phenomenon.

In the zeroth order of $g$, the long-time asymptotic solution of $\alpha (t)$ is given by
\beq
\alpha_{\rm 0}(t)=-\rmi \frac{\xi}{\kappa} e^{-\rmi \omega_{\rm e} t},\label{alpha0}
\eeq
where we use the fact that $\omega_{\rm p}=\omega_{\rm e}$, and $\omega_{\rm p}\gg \kappa$.
Substituting this solution into the MF Hamiltonian for an atom~(\ref{atom_Hamiltonian}),
we obtain the effective spin model for this order:
\begin{numcases}{\qquad\qquad\fl
}
{\cal H'}_{\rm spin (D)}=\sum_{j=1}^N \omega_{\rm a}S_j^z-\sum_{j=1}^N \frac{4g\xi}{\kappa} \sin (\omega_{\rm e} t) S_j^x &(Dicke),\label{eff_Dicke}\\
{\cal H'}_{\rm spin (TC)}=\sum_{j=1}^N \omega_{\rm a}S_j^z-\sum_{j=1}^N \frac{2g\xi}{\kappa}\left[ \sin (\omega_{\rm e}t) S_j^x - \cos (\omega_{\rm e} t) S_j^y \right] &(T-C).\label{eff_T-C}
\end{numcases}
As is clearly observed, the difference of the symmetry in both models appears in the different forms of the driving field.
That is, the driving field is polarized linearly in the $x$-direction for the Dicke model, while it gives a rotational field for the Tavis-Cummings model.
In order to understand the properties of the cavity system in the region of a strong driving field,
we first consider the Hamilton dynamics for the models~(\ref{eff_Dicke}) and (\ref{eff_T-C}) without a dissipation effect.

In the case of the Dicke model~(\ref{eff_Dicke}), it is known that this model shows a coherent destruction of tunneling (CDT)~\cite{Grifoni1998,Grifoni1991,Grossmann1991,Miyashita1998,Hijii2010} under a periodic driving field.
When CDT occurs, the state after one period of the external field does not change due to a quantum interference effect, and thus the atomic state is localized in the same state.
The condition for CDT is given by the zeros of the zero-order Bessel function~\cite{Gomez1992, Kayanuma1994},
\beq
J_0 \left( \frac{4g\xi}{\kappa\omega_{\rm e}} \right)=0.\label{CDT_zero}
\eeq
This relation shows that CDT repeatedly appear as a function of $\xi$.
In the phase diagram depicted in figure~\ref{diagram}(a), we plot these points by using blue curves.
We find a good agreement between the periodicity of the ordered phase and that of the curves for CDT, and thus we believe that CDT is the key ingredient for this symmetry-breaking phenomenon.

In contrast, the effective model for the Tavis-Cummings model~(\ref{eff_T-C}) shows Rabi oscillations
where the $xy$-component of the spin rotates with the frequency $\omega_{\rm e}$,
and thus there is no localization of the atomic state.
In this situation, the symmetry breaking in the $xy$-plane is not expected.

Thus, it is strongly suggested that the essence of the non-equilibrium phase transition is related to the CDT phenomenon.
However, if we take the dissipation effect into account in the model~(\ref{eff_Dicke}), the quantum dynamical motion originating from CDT is destroyed
and the stationary state is simply in the regular oscillating phase.
Thus, the leading-order term is not sufficient to understand the mechanism of the present phase transition.

The interaction among atoms comes from the next-to-leading-order term (see appendix B),
and the effective spin model is given by
\beq
{\cal H}_{\rm spin (D)}=\sum_{j=1}^N \omega_{\rm a}S_j^z-\sum_{j=1}^N \frac{4g\xi}{\kappa} \sin (\omega_{\rm e} t) S_j^x -\sum_{j,k}^N \frac{4 g^2}{N\omega_{\rm p}} S_j^x S_k^x.
\eeq
Here, the term of $\Or(g^2)$ describes the long-range interaction among atoms through the processes involving virtual emission and absorption of a photon.
This interaction effect, in cooperation with the CDT effect, should be crucial for the symmetry-breaking phenomenon found in the phase diagram of the Dicke model.
That is, we believe that the microscopic effect of quantum interference (CDT) is enhanced by the strong interaction among atoms,
and appears to be a macroscopic cooperative phenomenon.
    
\subsection{Stationary-state dependence on the dissipation constants}\label{bath_type}
\begin{figure}
\begin{center}
\begin{tabular}{cccc}
(a)&&(b)\\
&\includegraphics[width=64mm,keepaspectratio,clip]{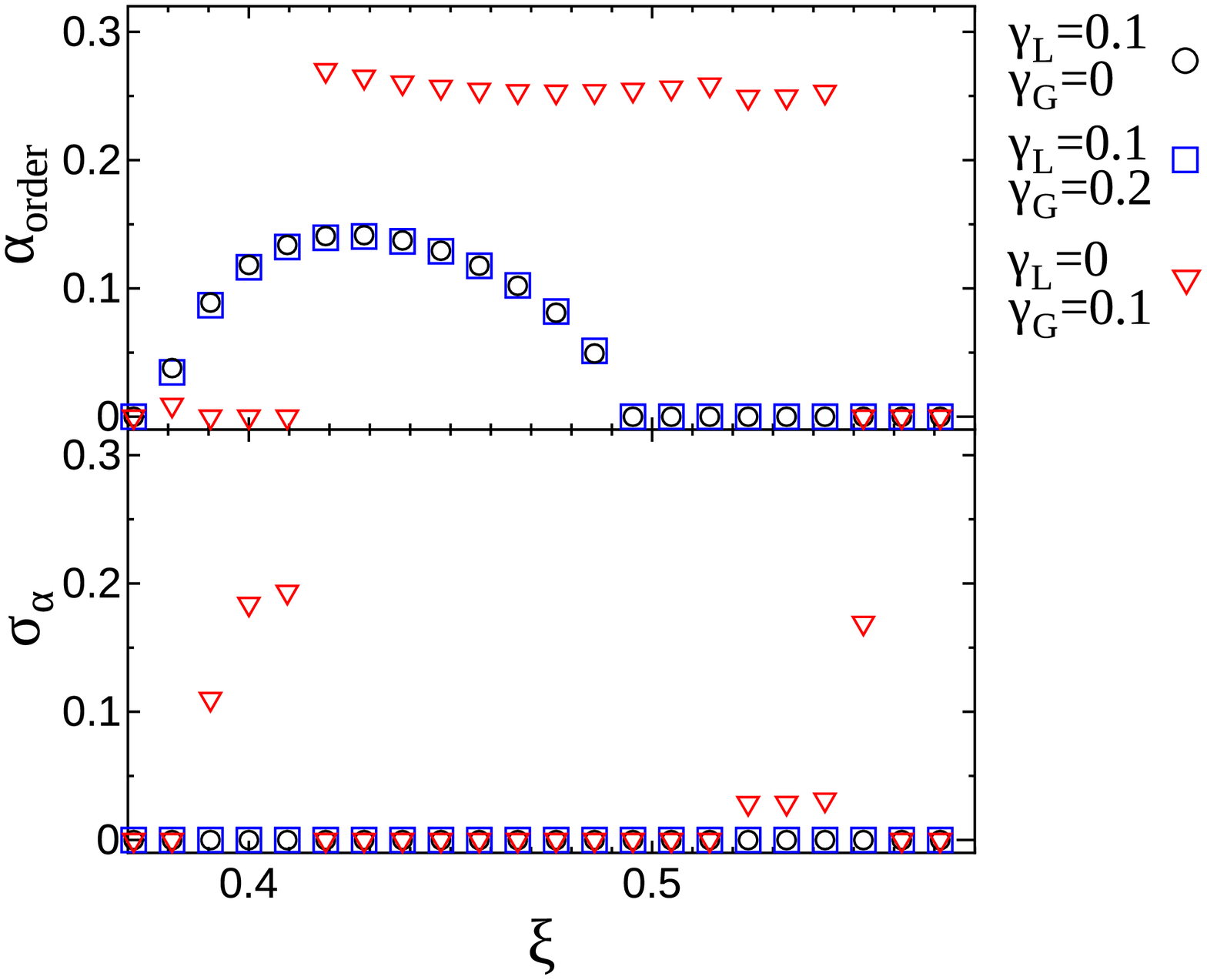}
&&\includegraphics[width=64mm,keepaspectratio,clip]{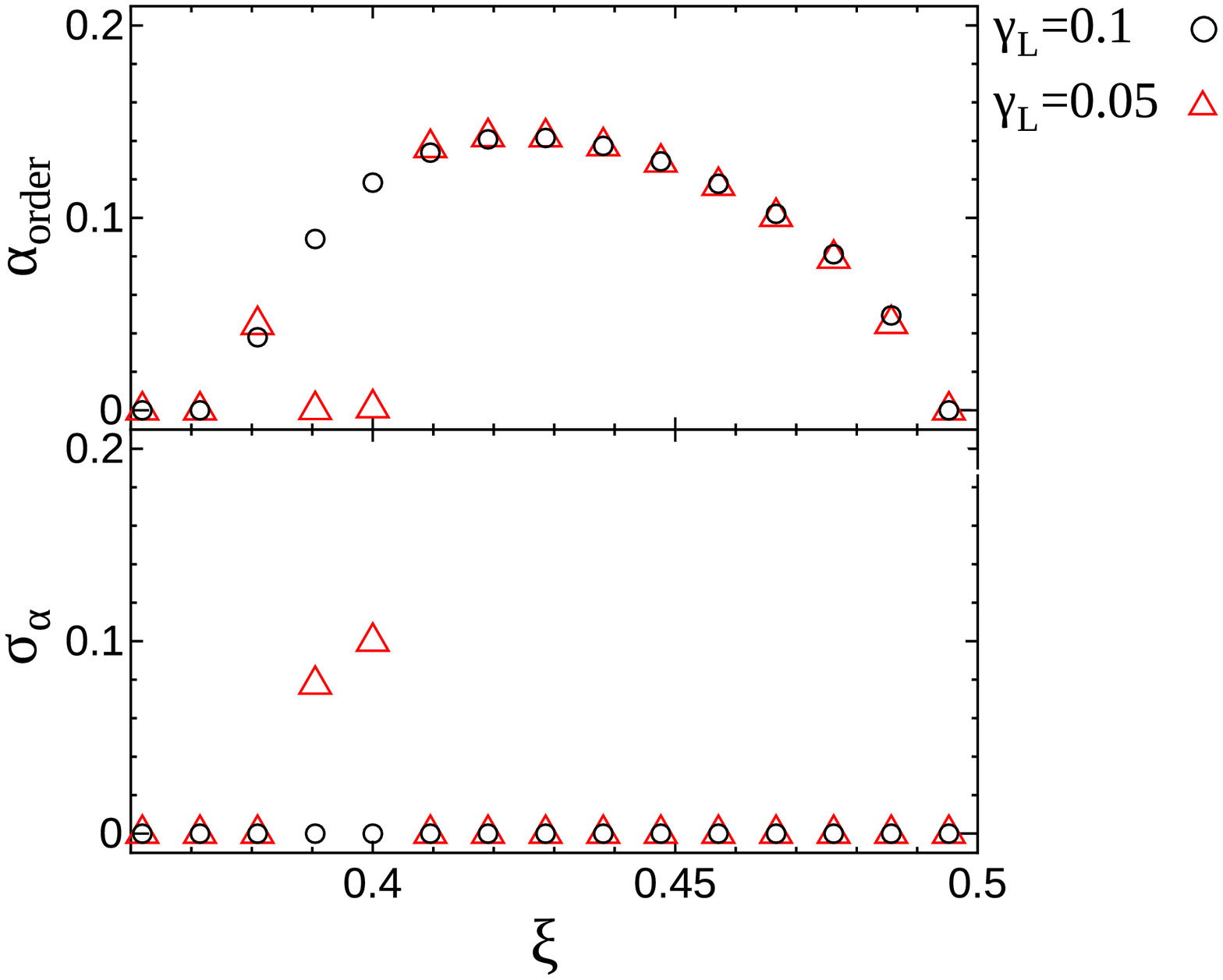}
\end{tabular}
\end{center}
\caption{
(a) Dependences of the photon order parameter $\alpha_{\rm order}$ (top) and $\sigma_{\alpha}$ (bottom) on the dissipation constant $\gamma_{\rm G}$ for the Dicke model.
Almost the same dependence is observed as long as $\gamma_{\rm L} \ne 0$.
For the case $\gamma_{\rm L}=0$ (inverted triangles),
a qualitatively different dependence is observed because the total spin is conserved.
The dissipation constant $\kappa$ is set to be $0.1$.
(b) Dependence of the photon order parameter $\alpha_{\rm order}$ (top) and $\sigma_{\alpha}$ (bottom) on the dissipation constant $\gamma_{\rm L}$ for the Dicke model.
The local coupling bath ($\gamma_{\rm L}$) stabilizes the ordered state.
The dissipation constants $\kappa$ and $\gamma_{\rm G}$ are set to be $0.1$ and $0$, respectively.
}
\label{dissipation_effect}
\end{figure}

In the above simulations, we adopted only the local coupling bath $\gamma_{\rm L}=0.1$.
Here, we discuss the dependence of $\alpha_{\rm order}$ and $\sigma_{\alpha}$ on the various types of thermal baths.

First, we study the effect of a global-coupling bath $\gamma_{\rm G}$ on the stationary states.
In figure~\ref{dissipation_effect} (a), we show the dependence for various sets of $\gamma_{\rm L}$ and $\gamma_{\rm G}$.
We plot circles for $\gamma_{\rm L} = 0.1$ and $\gamma_{\rm G} = 0$ (shown in figure~\ref{driven_order_Dicke}), squares for $\gamma_{\rm L} = 0.1$ and $\gamma_{\rm G} = 0.2$, and inverted triangles for $\gamma_{\rm L} = 0$ and $\gamma_{\rm G} = 0.1$.
When the dissipative environment consists of only the global coupling bath (inverted triangles), the total spin $(\sum_{j=1}^N \bi{S}_j)^2$ is conserved,
and we find a qualitatively different behaviour from other cases with non-zero $\gamma_{\rm L}$.
As long as the total spin is not conserved $(\gamma_{\rm L} \neq 0)$,
the stationary state does not depend considerably on $\gamma_{\rm G}$.
We also checked that the effect of the global coupling bath $(\gamma_{\rm G})$ does not cause a qualitative change in other parameter regions.

Furthermore, we study the dependence of the stationary states on the strength of the dissipation constant $\gamma_{\rm L}$.
In figure~\ref{dissipation_effect}(b), we show the data for $\gamma_{\rm L} = 0.1$ and $\gamma_{\rm G} = 0$ (circles) and for $\gamma_{\rm L} = 0.05$ and $\gamma_{\rm G} = 0$ (triangles).
When $\gamma_{\rm L}$ becomes small, 
in some values in which the system is in the ordered states for $\gamma_{\rm L}=0.1$, the states change to the non-periodic states.
This observation shows that $\gamma_{\rm L}$ stabilizes the ordered states.

We also checked the effect of the dissipation constant $\kappa$ by changing the value of $\kappa$ and $\xi$,
and found that the properties of the stationary states remain almost unchanged when the scaled parameter $\xi/\kappa$ is held constant (not shown).
Recalling that the conditions under which CDT occurs depend on the dimensionless parameter $g\xi/\kappa\omega_{\rm e}$~(\ref{CDT_zero}),
this observation is consistent with the physical interpretation that the dynamical effect of CDT is important for this non-equilibrium phase transition.

\section{Summary}\label{Sec_five}
In the present paper, we studied the cooperative phenomena of photons and atoms in a cavity under a driving external field.
In order to study the region where the interaction and the driving external field are strong,
we derived the dressed Lindblad equations (\ref{extend_photon}) and (\ref{extend_atom}), which incorporate the atom-photon coupling and the driving external field into the dissipation terms.
In this derivation, we made use of the fact that a mean-field treatment is exact, thanks to the uniform coupling between photons and many atoms.

Applying the derived dressed Lindblad equation, we found a novel symmetry-broken state appearing repeatedly in the phase diagram for the Dicke model (figure~\ref{diagram}(a)).
In this state, the components of photon field and atomic excitation driven by the external field exhibit a symmetry-breaking phenomenon.
For a physical interpretation of this phenomenon,
we discussed an effective spin model,
and we concluded that the phenomenon originates from the microscopic quantum interference effect (CDT)
enhanced by the interaction among atoms induced by the uniformly coupled cavity photons.

\ack
The authors thank Dr Patrice Bertet, Professor Irinel Chiorescu and Professor Hans De Raedt for their valuable comments and discussion,
and thank Dr Sergio Andraus for carefully reading the paper.
The present work was partially supported by the Mitsubishi Foundation and also the Next Generation Super Computer Project (Nanoscience Program from MEXT of Japan).
TS, TM and SM acknowledge the support from JSPS grant no. 258794, grant no. 227835, and grant no. 25400391, respectively.
The numerical calculations were supported by the supercomputer center of ISSP of the University of Tokyo.
This work was supported by the Sumitomo Foundation.

\appendix

\section{Dicke transition in the bare Lindblad equation}\label{AppendixA}
\begin{figure}
\begin{center}
\includegraphics[width=50mm,keepaspectratio,clip]{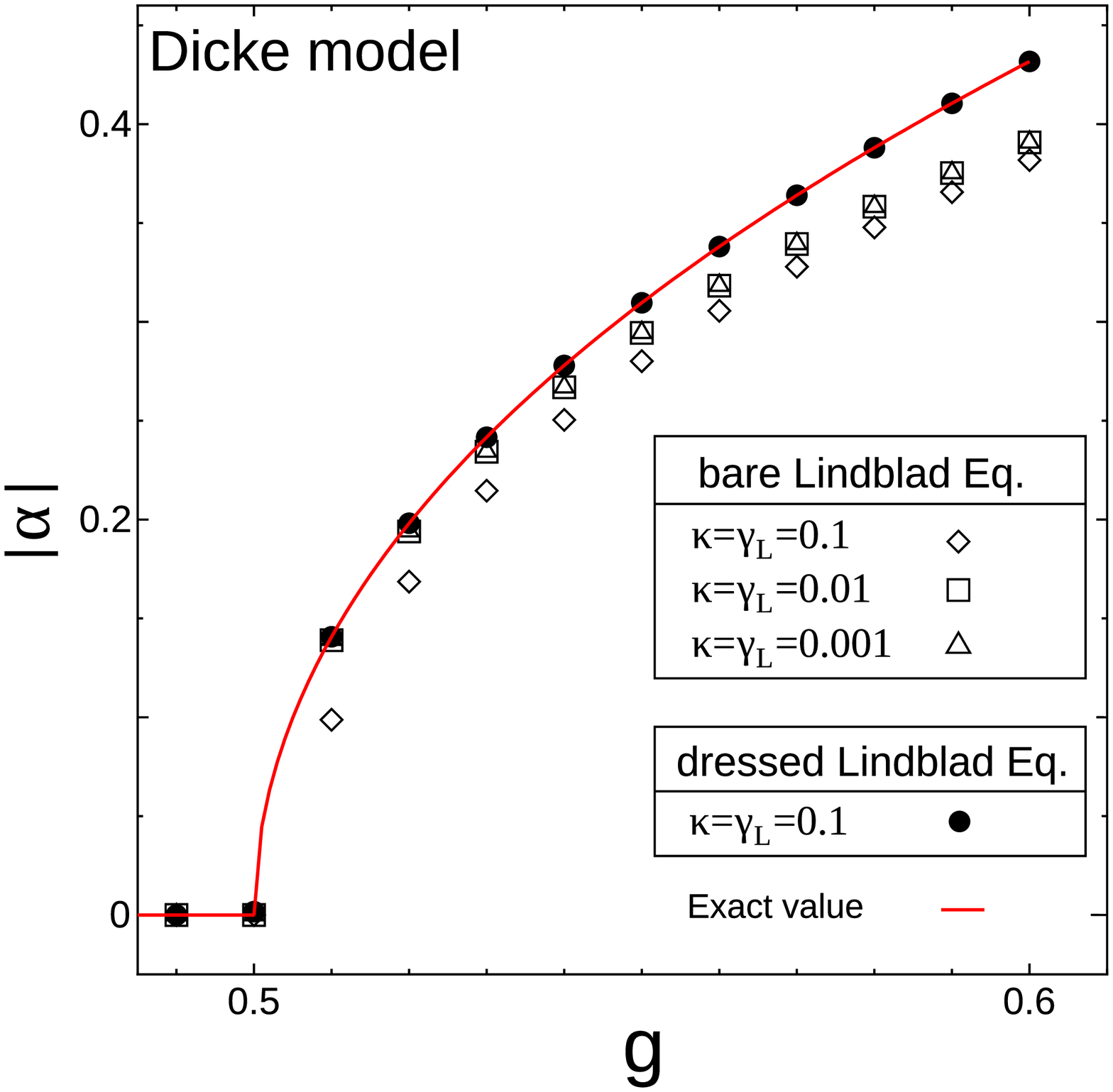}
\end{center}
\caption{
Ordered component of the photon field $\left| \alpha \right|=\left| \braket{a}/\sqrt{N} \right|$ for stationary states in the Dicke model as a function of the interaction strength $g$.
Diamonds, squares, and triangles give the data obtained by the bare Lindblad equations~(\ref{Lindblad_photon}) and (\ref{Lindblad_atom})
with $\kappa=\gamma_{\rm L}=0.1, 0.01$ and $0.001$, respectively.
They converge to a limiting value as $\kappa$ and $\gamma_{\rm L}$ decrease, but this value is not correct.
Bullets show results obtained by the dressed Lindblad equations~(\ref{extend_photon}) and (\ref{extend_atom}), which agree with the exact results given by the red curve $\left| \alpha \right|=\frac{1}{2} [4g^2/\omega_{\rm p}^2 -\omega_{\rm a}^2/(4g^2)]^{1/2}$~\cite{Hepp1973}.
The dissipation constant $\gamma_{\rm G}$ is set to be zero.
}
\label{no_Dicke}
\end{figure}

In this appendix, we demonstrate that the bare Lindblad equations,
\begin{numcases}{}
\frac{\rmd \rho_{\rm p}(t)}{\rmd t}=-\rmi [H_{\rm p}(t), \rho_{\rm p}(t)]+\kappa (2 a \rho_{\rm p} (t) a^{\dagger}-\{ a^{\dagger} a, \rho_{\rm p}(t) \} ),\label{Lindblad_photon}\\
\frac{d \rho_{\rm a}(t)}{dt}=-\rmi [H_{\rm a} (t), \rho_{\rm a}(t)]+\gamma_{\rm L} (2 S^- \rho_{\rm a} (t) S^+ -\{ S^+ S^-, \rho_{\rm a}(t) \} )\nonumber\\
\qquad\qquad+\gamma_{\rm G} (\langle S^- \rangle [ \rho_{\rm a} (t), S^+] + \langle S^+ \rangle [S^-, \rho_{\rm a} (t) ] ),
\label{Lindblad_atom}
\end{numcases}
do not give the correct ground state for the Dicke transition at $\xi=0$.
Since the dissipation terms in (\ref{Lindblad_photon}) and (\ref{Lindblad_atom}) do not take the effect of the interaction into account,
the applicability of this master equation is expected to be limited to the region where the atom-photon interaction is weak.
In figure~\ref{no_Dicke},
we plot the dependence of the ordered component of the photon field $|\alpha|=\lim_{N \rightarrow \infty}|\braket{a}/\sqrt{N}|$ in the stationary states for the Dicke model.
In the figure, we show the stationary states obtained by the bare Lindblad equations for $\kappa=\gamma_{\rm L}=0.1$ (diamonds), $\kappa=\gamma_{\rm L}=0.01$ (squares), and $\kappa=\gamma_{\rm L}=0.001$ (triangles).
The bare Lindblad equations~(\ref{Lindblad_photon}) and (\ref{Lindblad_atom}) give a phase transition,
but fail to give the exact values of the order parameter.
This deviation is not due to the finite dissipation constants, $\kappa$ and $\gamma_{\rm L}$,
but due to the form of the master equation where the effect of the interaction is not incorporated.
Indeed, as $\kappa$ and $\gamma_{\rm L}$ decrease, the data converge to limiting values which are deviated from the correct values.
For the Tavis-Cummings model, the bare Lindblad equations do not reproduce the Dicke transition at all (not shown).
This indicates that the bare Lindblad equations are inadequate for studying the USC region.
On the other hand, the dressed Lindblad equations~(\ref{extend_photon}) and (\ref{extend_atom}) reproduce the exact values (red line) for the Dicke model (bullets) and also for the Tavis-Cummings model (not shown).

\section{Long-range interaction among atoms in the effective spin model}\label{AppendixB}
In this appendix, we derive the next-to-leading-order term of the effective spin model which describes the long-range interaction. 
In the equation of motion of the photons~(\ref{photon_eq_Dicke}), $\alpha_1 (t)$ which is the order of $g$ obeys
\beq
\frac{d}{dt}\alpha_1 (t)=(-\rmi \omega_{\rm p}-\kappa )\left( \alpha_1 (t)+\frac{2g}{\omega_{\rm p}} m^x (t) \right).
\eeq
The long-time asymptotic solution of $\alpha_1 (t)$ is given by
\beq
\alpha_1 (t)=(-\rmi \omega_{\rm p}-\kappa )\frac{2g}{\omega_{\rm p}}\int_0^{\infty} m^x(t-t')e^{(-\rmi \omega_{\rm p}-\kappa)t'} dt'.\label{alpha_first}
\eeq
The dominant contribution of $m^x (t-t')$ to the integral comes from the component oscillating around the frequency $\omega_{\rm p}$ within the time window $0<t'<1/\kappa$.
Therefore, we focus on two time scales the short time scale $\Or (1/\omega_{\rm p})$ and the long time scale $\Or (1/\kappa)$.
In the present simulation, the relation $\kappa \gg (\gamma_{\rm L}(t), \gamma_{\rm G}(t))$ holds because the eigenstate of the dressed atom is almost the same as that of $S^x$ for most of the time under a strong driving field,
and therefore we find, from the definition of $\gamma_{\rm L}(t)$~(\ref{local_coupling_bath}) and $\gamma_{\rm G}(t)$~(\ref{global_coupling_bath}), that the dissipation strength is strongly suppressed.
Therefore, the atomic motion within this time window $0<t'<1/\kappa$ is described by the Hamilton dynamics:
\begin{numcases}{\qquad\qquad\fl
}
\frac{d}{dt}m^x (t)=-\omega_{\rm a} m^y (t),\label{m^x}\\
\frac{d}{dt}m^y (t)=\omega_{\rm a} m^x (t)-2g (\alpha (t)+\alpha^* (t)) m^z (t) \simeq \omega_{\rm a} m^x (t)+\frac{4g\xi}{\kappa} \sin (\omega_{\rm e} t) m^z (t),\\
\frac{d}{dt}m^z (t)= 2g (\alpha (t)+\alpha^* (t)) m^y (t)\simeq -\frac{4g\xi}{\kappa} \sin (\omega_{\rm e} t) m^y (t),
\end{numcases}
where we substitute $\alpha_0 (t)$ into $\alpha (t)$.
In the short time scale $\Or (1/\omega_{\rm p})$, since $m^y (t)$ oscillates many times in the time interval of $1/\omega_{\rm p}$ under a strong driving field ($\xi \gg \kappa \omega_{\rm p}/g$), 
the RHS of equation~(\ref{m^x}) is regarded as being zero for this time scale.
On the other hand, in the long time scale $\Or (1/\kappa)$, $m^x (t)$ oscillates with frequency $\omega_{\rm a} J_0 (4g\xi/\kappa\omega_{\rm e})$~\cite{Kayanuma1994}.
Hence, when the driving field is sufficiently strong, i.e., $\omega_{\rm a}J_0 (4g\xi/\kappa\omega_{\rm e}) \ll \kappa$,
we regard $m^x (t-t')$ as a constant also for the long time scale. 
Thus, under the strong driving field, we evaluate
\beq
m^x (t-t')=m^x (t) \quad {\rm for} \quad 0<t'<1/\kappa,
\eeq
and therefore from equation~(\ref{alpha_first}), we have
\beq
\alpha_1 (t)=-\frac{2g}{\omega_{\rm p}} m^x (t).\label{alpha1}
\eeq
Substituting the long-time asymptotic solution of the photons up to the first order of $g$, equations~(\ref{alpha0}) and (\ref{alpha1}), into~(\ref{atom_Hamiltonian}),
we obtain the following effective spin model,
\beq
{\cal H}_{\rm spin (D)}=\sum_{j=1}^N \omega_{\rm a}S_j^z-\sum_{j=1}^N \frac{4g\xi}{\kappa} \sin (\omega_{\rm e} t) S_j^x -\sum_{j,k}^N \frac{4 g^2}{N\omega_{\rm p}} S_j^x S_k^x.
\eeq
The term of the second order in $g$ describes the long-range interaction among atoms.

\end{document}